\documentclass[doublecol]{epl2}
\usepackage{graphicx,macros,amsmath,amssymb}

\title{Escalation of error catastrophe for enzymatic self-replicators}
\author{Benedikt Obermayer \and Erwin Frey\thanks{E-mail: \email{frey@physik.lmu.de}}} 
\shortauthor{Obermayer et al.}
\institute{Arnold-Sommerfeld-Center for Theoretical Physics and Center for NanoScience, Ludwig-Maximilians-Universit\"at M\"unchen, Theresienstr. 37, 80333 M\"unchen, Germany}

\abstract{It is a long-standing question in origin-of-life research whether the information content of replicating molecules can be maintained in the presence of replication errors. Extending standard quasispecies models of non-enzymatic replication, we analyze highly specific enzymatic self-replication mediated through an otherwise neutral recognition region, which leads to frequency-dependent replication rates. We find a significant reduction of the maximally tolerable error rate, because the replication rate of the fittest molecules decreases with the fraction of functional enzymes. Our analysis is extended to hypercyclic couplings as an example for catalytic networks.}


\begin{document}

\maketitle

\section{Introduction} 

According to the RNA world hypothesis~\cite{OrgelCRBM:04}, prebiotic biochemical life is thought to have emerged through four steps: starting from the primordial non-enzymatic synthesis of nucleotides and their subsequent non-enzymatic polymerization into random RNA, which in a third step would non-enzymatically replicate, natural selection would finally produce a set of functional RNA enzymes (ribozymes), establishing exponential growth and initiating RNA evolution. Despite considerable experimental progress~\cite{JohnstonScience:01,LincolnScience:09}, as of today no truely self-replicating system has been evolved according to this hypothetic schedule. To assess its intrinsic
plausibility, theory has mainly focused on the third step, usually  based on the Eigen model~\cite{EigenACP:89} for prebiotic evolution: here, auto-catalytic self-replication of $L$-nucleotide sequences proceeds non-enzymatically via stepwise template-directed polymerization, with a non-negligible error probability $\mu$ per single nucleotide. Assuming that one specific ``master'' template replicates with the highest rate $\alpha > 1$, while all other sequences have unit replication rate, it is found that faithful replication of the master is possible only for error probabilities smaller than a critical value $\muc\approx \ln \alpha/L$. In this regime, the population in sequence space is concentrated about the master in a rather broad distribution, giving rise to the notion of a ``quasispecies''. Larger values $\mu > \muc$  lead to a delocalized state with completely random sequences in the population. Many aspects of the Eigen model depend to a large extent on the chosen fitness landscape, which assigns replication rates to genotypes. In the case of RNA, it displays a considerable degree of neutrality, because the mapping of sequences to secondary structures is decidedly many-to-one~\cite{HuynenPNAS:96}. Still, although not universal, the existence of a critical mutation rate $\muc$ is a comparatively robust phenomenon~\cite{WieheGR:97,Jain:05}.  It has been termed ``error catastrophe''~\cite{EigenNaturwissenschaften:78}, because it puts possibly irreconcilable simultaneous constraints on maximally tolerable error probability and minimal functional sequence length.

Lacking actual observations of freely self-replicating RNA and hence reliable estimates for replication rates, these theoretical limitations of non-enzymatic RNA replication are not yet reasonably quantitative. However, biochemical issues~\cite{OrgelNature:92} raise severe doubts about its plausibility as well. Although ribozymes have been discovered that catalyze most of the necessary reaction steps~\cite{LincolnScience:09,JohnstonScience:01,JoyceAC:07,LilleyCOSB:05}, it remains questionable how a ribozyme should literally copy itself~\cite{JoyceAC:07,SzostakNature:01}. Enzymatic replication seems the far more likely scenario, in the sense that a ribozyme copies other molecules. Presumably and most effectively, it would copy only those molecules that are exact replicas of itself, not only because known ribozymes act very substrate-specific, but also because unspecific recognition does not give a selective advantage to the replication enzymes themselves; this would require compartmentalization in vesicles to keep closely related molecules together~\cite{SzostakNature:01}. Further, it has recently been suggested that the spontaneous emergence of RNA polymerases even without previous non-enzymatic replication could be promoted by a significant increase of functional complexity in a pool of random RNA due to the likely appearance of ligase activity~\cite{BrionesRNA:09}.

In order to comparatively analyze non-enzymatic and enzymatic replication, their competition and their respective tolerance against mutations theoretically and by means of stochastic simulations, we employ a simplified quasispecies model, where sequences replicate both non-enzymatically and enzymatically, the latter with high specificity. We find a coexistence regime of these two replication modes, and an escalation of the error catastrophe in the enzymatic case: because the replication rate of the fittest molecules decreases with the fraction of functional enzymes, the maximally tolerable mutation rate is significantly reduced. To make contact to models of modular evolution and catalytic networks, where complex function is assumed to emerge through independent selection of small functional motives, thereby circumventing the error catastrophe~\cite{ManrubiaRNA:07,TakeuchiBD:08}, we then extend our analysis to the case of hypercycles~\cite{EigenNaturwissenschaften:78,CamposPRE:00,SilvestreJTB:08}.

\begin{figure}
   \centerline{\includegraphics[width=.48\textwidth]{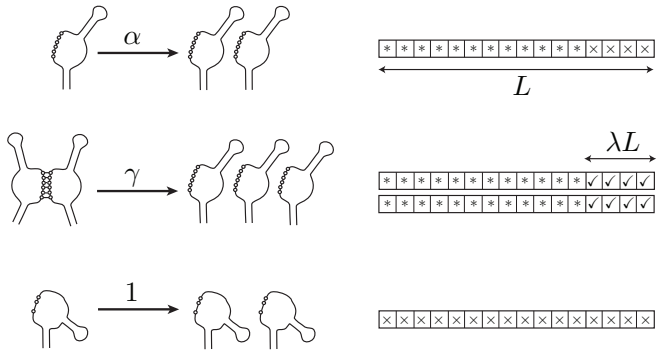} }
  \caption{\label{fig:model}Schematic illustration of the model. Left:  Molecules with correct structure can replicate non-enzymatically with rate $\alpha > 1$. These molecules can also replicate enzymatically with rate $\gamma$, if they bind specifically to an identical partner within an otherwise selectively neutral recognition region of $\lambda L$ sites (dots). Misfolding mutant molecules replicate with unit rate. This model is formulated in terms of sequences rather than structures, as shown on the right panel: we distinguish correct ``structural'' nucleotides ($*$), matching sites in the recognition region ($\checkmark$) and unmatching or random nucleotides ($\times$).}
\end{figure}

\section{Model}
Motivated by the observation that catalytic and recognition regions
are often clearly separated in ribozymes like the RNA component of
RNaseP~\cite{LilleyCOSB:05}, we assume that the specific recognition mediating enzymatic replication involves only a small fraction $\lambda$ of otherwise selectively neutral sites. This means that the majority of sites forms the proper secondary structure of the molecule and builds its active center, which catalyzes the polymerization reactions. Although secondary structure folding algorithms provide an improved genotype-fitness mapping through an excellent approximation to RNA phenotypes, our model is formulated in terms of sequences instead of structures to allow for analytical treatment. We hence distinguish between ``structural sites'' and a ``recognition region'' on the sequence level (see Fig.~\ref{fig:model} for a schematic illustration of our model).  For the former, we use a sharply-peaked fitness landscape: a master sequence $S^*$ has the highest non-enzymatic replication rate $\alpha > 1$, while all other sequences replicate with unit rate defining the time scale. We ignore possibly neutral sites in the structural region, because on our level of approximations this merely renormalizes their total number, or, equivalently, the mutation probability (see below). However, we do account for mutations in the recognition region, which do not affect non-enzymatic replication but the specificity of enzymatic replication: idealizing ``highly specific'', we require the recognition regions of enzyme and substrate to be identical for enzymatic replication to take place. Hence, ribozymes replicate only exact copies of themselves, with $\gamma$ the associated rate constant. Note that we do not make any restrictions on the specific sequence of the recognition region: any molecule with the correct sequence for the structural sites can replicate enzymatically if it recognizes a suitable enzyme.

In the following, we formalize this model in the framework of
quasispecies theory~\cite{EigenACP:89}, where molecules are
represented by sequences $S_i=(\sigma^{(i)}_1\sigma^{(i)}_2\ldots
\sigma_L^{(i)})$ of $L$ binary nucleotides $\sigma_j\in\lbrace
0,1\rbrace$. Their concentrations $x_i$ evolve in the
$L$-dimensional hypercube according to the deterministic rate
equations
\begin{equation}\label{eq:rate-equations}
  \dot x_i = \sum_k m_{ik} r_k x_k - x_i\,\bar r,
\end{equation}
where $r_k$ is the replication rate of $S_k$,
$m_{ik}=\mu^{d_{ik}}(1-\mu)^{L-d_{ik}}$ is the mutation probability
between sequences $S_i$ and $S_k$ with Hamming distance $d_{ik}$, and
$\mu$ is the single-nucleotide mutation probability. The second term in Eq.~\eqref{eq:rate-equations} involves the mean replication rate
$\bar r=\sum_k r_k x_k$ and ensures the normalization $\sum_k
x_k=1$.  According to the above defined model, the
replication rates read
\begin{equation}\label{eq:replication-rate}
r_k = \begin{cases} \alpha + \gamma x_k, &\text{if } S_k
  \rvert_\text{struc}=S^*\rvert_\text{struc},\\
 1, &\text{otherwise.}
\end{cases}
\end{equation}
In Eq.~\eqref{eq:replication-rate}, $S_k \rvert_\text{struc}$ denotes the restriction of the sequence $S_k$ to the structural sites, and
$S^*\rvert_\text{struc}$ is the corresponding master sequence. While  
replication rates are usually taken as functions only of the genotype, with one single peak at the master sequence~\cite{EigenACP:89,SchusterBMB:88,NowakJTB:89,Jain:05,WieheGR:97,SaakianPNAS:06}, our model leads to frequency-dependent selection, which has only rarely been analyzed because it leads to mathematically challenging replicator-mutator equations (see, e.g., Ref.~\cite{StadlerBMB:95}).

\section{Stochastic simulation}
\begin{figure}
\includegraphics[angle=90,width=.48\textwidth]{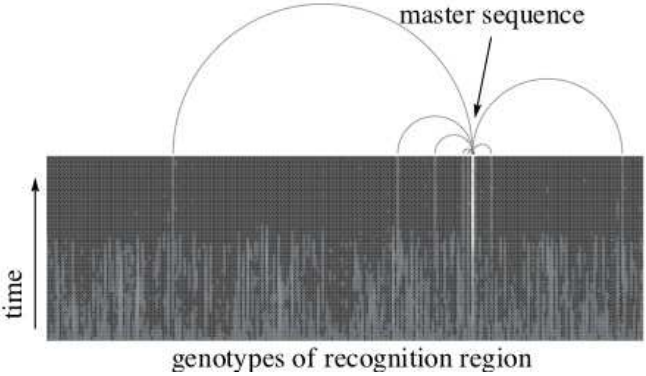}
\caption{\label{fig:timeseries} Exemplary run of the stochastic simulation in a population of $N=10^3$ sequences for $L=32$, $\alpha=5$, $\gamma=10$, $\lambda=1/4$, and $\mu=0.005$. All sequences have been initialized with correct structural region but random recognition region. Their concentration is shown in gray level as function of time and genotype in the recognition region (linearly arranged by reading bit strings as integer numbers). Spontaneous concentration fluctuations lead to the establishment of a quasispecies of enzymatic replicators centered about one specific yet randomly chosen master sequence. Neighboring sequences are indicated by thin lines.}
\end{figure}

For a realization of the full $2^L$-dimensional system Eq.~\eqref{eq:rate-equations} in a finite population of $N$ sequences, we employ the straightforward stochastic simulation algorithm used in Ref.~\cite{WilkePR:01}. At each time $t$ each sequence $S_k$, present in $n_k$ copies, has a probability $p_{0,k} = n_k/\sum_i n_i(1+r_i)$ to be copied without mutations into the population at time $t+1$, and a probability $p_{\text{mut},jk} = m_{jk} r_k n_k/\sum_i n_i (1+r_i)$ to be selected and mutated into sequence $S_j$. The population is initialized uniformly at the master sequence, but with random recognition sequences. Because the sites of the recognition sequence are effectively neutral, a state with all possible recognition sequences present in equal concentration is stable for large sequence length~\cite{StadlerBMB:95}. But if number fluctuations sufficiently increase the concentration of one particular \emph{yet randomly chosen} sequence, this conveys via Eq.~\eqref{eq:replication-rate} a selective advantage, and its concentration will thus increase, up to the extent that mainly this sequence and its next mutational neighbors are present, in a quasispecies distribution very much like the one obtained in usual fitness landscapes.  Fig.~\ref{fig:timeseries} shows an example of this outcome, which is somewhat reminiscent of a fixation event. While its detailed dependence on the specific formulation of the underlying stochastic process and the parameter values is left for future research, the localization itself turns out to be a robust phenomenon. In the following, we will therefore without loss of generality assume that the recognition region of the most populated sequence is equal to the one of the master sequence $S^*$.

\section{Results}
\begin{figure}
\centerline{\includegraphics[width=.48\textwidth]{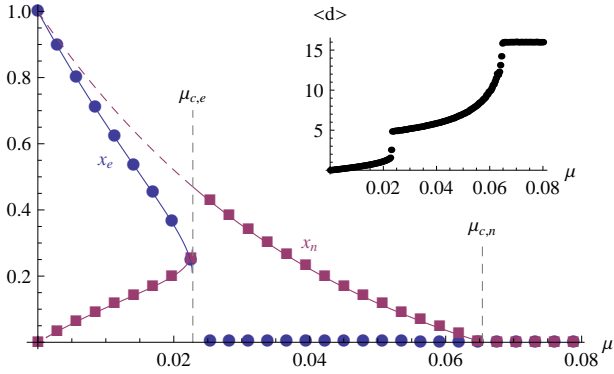}}
\caption{\label{fig:eigen-cat-plot}Comparison between simulation  results for $\xe$ (circles) and $\xn$ (squares) in a population of    $N=10^4$ sequences and solutions to    Eq.~\eqref{eq:rate-equations-error-tail} for $L=32$, $\alpha=5$, $\gamma=10$ and $\lambda=1/4$. The two error catastrophes occur at  $\mu_\text{c,e} \approx\ln\Qc^{-1}/L$ with $\Qc$ a solution of Eq.~\eqref{eq:discriminand} and $\mu_\text{c,n}\approx\ln\alpha/(L(1-\lambda))$.  The inset shows the average Hamming distance $\ave{d}$ to the master sequence, which increases in two steps, the first one at $\mu_\text{c,e}$ discontinuous, the second one at $\mu_\text{c,n}$ continuous.}
\end{figure}

While analytic solutions to the $2^L$-dimensional system
Eq.~\eqref{eq:rate-equations} are hard to obtain, we can use the
so-called ``error-tail'' approximation~\cite{NowakJTB:89}: here, we
introduce three different classes of molecules. In $\xe$, we gather enzymatic replicators identical to the master sequence, with a replication rate $\re=\alpha+\gamma \xe$. We use a second class
$\xn$ for non-enzymatic replicators, with structural sites identical to the master sequence but random recognition sequences. Their 
replication rate is $\rn=\alpha$: although they are capable of
enzymatic replication, the fraction of suitable enzymes with the
appropriate recognition region is negligible. Finally, $1-\xe-\xn$ is the error-tail of molecules with incorrect structural sites and unit replication rate. The main approximation of the error-tail approximation is to consider only those mutations that lead into a less-fitter class, with the probability not to have such a mutation abbreviated as ``quality factor'' $Q$. This approximation is generally valid for large sequence length but may fail if peaks in the fitness landscape are very dense~\cite{Jain:05}. The enzymatic replicators in $\xe$ have $\Qe=(1-\mu)^L\equiv Q$, because a single error in $L$ nucleotides suffices to destroy either structural or recognition region. The non-enzymatic replicators in $\xn$ have a larger quality factor $\Qn=(1-\mu(1-\lambda))^L\approx Q^{1-\lambda} > Q$: because the presence of $\lambda L$ neutral sites in the recognition region reduces the effective mutation probability, these sequences are mutationally more robust~\cite{WilkeNature:01,TakeuchiBEB:05,KunNG:05}. Further,
with probability $Q^{1-\lambda}-Q$ mutations  in $\xe$ will hit a site
of the recognition region and thus contribute to $\xn$. Hence, the
dynamical system in the error-tail approximation is given by
\begin{equation}
\label{eq:rate-equations-error-tail}
\begin{split}
\dot x_\text{e} &= \re Q \xe - \xe \bar r \\
\dot x_\text{n} &= \rn Q^{1-\lambda} \xn + \re (Q^{1-\lambda}-Q)\xe - \xn \bar r,
\end{split}
\end{equation}
where the mean replication rate reads $\bar r=
(\re-1)\xe+(\rn-1)\xn+1$. Solutions to the stationary state $\dot x_\text{e}=\dot x_\text{n}=0$ of Eq.~\eqref{eq:rate-equations-error-tail} for different mutation probabilities $\mu$ are shown in Fig.~\ref{fig:eigen-cat-plot}, together with results from a stochastic simulation of the full system with the replication rates Eq.~\eqref{eq:replication-rate} in a population of $N=10^4$ sequences, where we initialized the sequences uniformly at the master sequence to reduce noise resulting from the intrinsically stochastic ``fixation'' events shown in Fig.~\ref{fig:timeseries}, and averaged the results over time after reaching a stationary state. Obviously, approximating the deterministic rate equations with the simplified Eq.~\eqref{eq:rate-equations-error-tail} gives an excellent description of the stochastic system. We can clearly distinguish three different regimes, separated by two error thresholds.

For high mutation probability, the population is delocalized over sequence space and only the error tail is significantly populated ($\xe=\xn=0$). For smaller values of $\mu$, we find a ``non-enzymatic regime'', where sequences with correct structural region are present, but a stable recognition sequence cannot be maintained, such that enzymatic replication is not possible. Explicitly, we find $\xe=0$ and $\xn=(\alpha Q^{1-\lambda}-1)/(\alpha-1)$. The two regimes exchange stability at $Q=\alpha^{-1/(1-\lambda)}$, corresponding to $\mu=\mu_\text{c,n} \approx \ln \alpha/(L(1-\lambda))$. This is the familiar ``phenotypic error threshold''~\cite{KunNG:05,TakeuchiBEB:05}: the presence of neutral sites renormalizes the effective mutation probability, equivalent to having a shorter sequence~\cite{EigenACP:89}. 

For smaller mutation probabilities $\mu < \mu_\text{c,e}$ the ``enzymatic regime'' becomes stable. Here, the fraction $\xe$ of enzymatic replicators is nonzero, but $\xn > 0$ as well, because this class is fed from  $\xe$ through mutations in the recognition region. Solving a third-order polynomial for $\xe$ and $\xn$, we find this regime is stable when the corresponding discriminand is positive, which yields a critical value $Q=\Qc$ from the condition
\begin{multline}\label{eq:discriminand}
  4\left[3\gam(1+\alpha\Qc(\Qc^{-\lambda}-2))-(\gam\Qc+\Qc^{-\lambda}-\alpha)^2\right]^3\\  
  +\left[9\gam(\alpha-\gam\Qc-\Qc^{-\lambda})(1+\alpha\Qc(\Qc^{-\lambda}-2))\right.\\ 
    +27\alpha\gam(\alpha\Qc-1)(1-\Qc^{-\lambda})
    \left.+2(\gam\Qc+\Qc^{-\lambda}-\alpha)^3\right]^2=0.
\end{multline}
Asymptotic solutions are given by:
\begin{equation}
  \label{eq:spl-plateau-Qc-asymptotes}
  \Qc\sim\begin{cases}
    1-\frac{\gam}{4\alpha\lambda},&\quad\text{if }\gam\ll\alpha,\\
    \frac{2\sqrt{\gam}-1}{\gam}
    +\frac{\lambda}{\gam}\ln\frac{2\sqrt{\gam}-1}{\gam} +     \mathcal{O}(\lambda^{2}), &\quad\text{if }\gam\gg\alpha.
  \end{cases}
\end{equation}
Note that the large-$\gamma$-limit is $\Qc\sim 2/\sqrt{\gamma}$, which implies for the corresponding critical value $\mu_\text{c,e}\approx \ln \Qc^{-1}/L\approx \ln\gamma/(2 L)$.  This significant reduction by a factor of 2 can be phrased as ``escalation of error catastrophe'': as the fraction $\xe$ of enzymatically replicating sequences drops with higher mutation probability, their replication rate $\re=\alpha+\gamma\xe$ decreases as well, leading to an even stronger reduction in $\xe$. Beyond the critical value $\mu_\text{c,e}$, the fraction of molecules with the correct recognition sequence becomes so small that their replication rate is not large enough for them to be maintained in the population at a macroscopic level.

\begin{figure}
\centerline{\includegraphics[width=.48\textwidth]{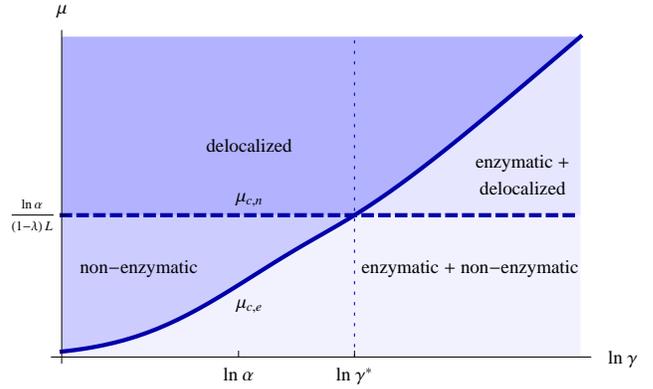}}
\caption{\label{fig:phase-diagram-enzymatic}Phase diagram of stability regimes of Eq.~\eqref{eq:rate-equations-error-tail} in the $\ln\gamma$-$\mu$-plane: the critical value $\mu_\text{c,n}$ (thick dashed line) separates the delocalized regime (above) and the non-enzymatic regime (below). Enzymatic replication is stable below $\mu_\text{c,e}$ (thick solid line) and becomes mutationally more robust than non-enzymatic replication if $\gamma > \gamma^*=\mathcal{O}(\alpha^2)$ (vertical line).}
\end{figure}

An important difference between the transitions at $\mu_\text{c,n}$ and $\mu_\text{c,e}$ can be observed not only in the fraction $\xe$, but also in the width of the population distribution (measured as average Hamming distance to the master sequence), shown in the inset of Fig.~\ref{fig:eigen-cat-plot}: while the delocalization transition at $\mu=\mu_\text{c,n}$ is continuous, the transition at $\mu_\text{c,e}$ is discontinuous. In the former case, this property depends also on the choice of observable~\cite{Jain:05}, but in the latter case, the discontinuity results from bistability: together with the  enzymatic regime, also the non-enzymatic regime or the delocalized phase may be stable, depending on whether $\mu$ is larger or smaller than $\mu_\text{c,n}$, and if $\mu > \mu_\text{c,e}$ the enzymatic regime vanishes. The phase diagram in Fig.~\ref{fig:phase-diagram-enzymatic} summarizes these various regimes. Note that $\mu_\text{c,e}=\mu_\text{c,n}$ at a critical value
\begin{equation}\label{eq:critical-gamma}
\gamma^* = \alpha(\alpha-1) + 2\lambda^{1/2}\alpha\sqrt{2\alpha(\alpha-1)\ln \alpha}+\mathcal{O}(\lambda).
\end{equation}
This result implies that very large rates $\gamma=\mathcal{O}(\alpha^2)$ are required if enzymatic replication is to be more error-tolerant than non-enzymatic replication~\cite{CamposPRE:00}. Although this possibility is not contained in the approximate Eq.~\eqref{eq:rate-equations-error-tail}, we find that in the bistability  region $\mu < \min (\mu_{\text{c,e}}, \mu_{\text{c,n}})$ the enzymatic regime is easily populated by selectively advantageous concentration fluctuations from the non-enzymatic regime by randomly choosing a ``master'' sequence for the recognition region as shown in Fig.~\ref{fig:timeseries}.

\section{Extension to hypercyclic couplings}
The realization that replication errors limit the maximum complexity of self-replicating molecules to a possibly paradoxical extent has lead to theories of \emph{modular} evolution, where complex functions emerge through catalytic interactions of smaller independently selected motifs~\cite{ManrubiaRNA:07,TakeuchiBD:08}, thereby also speeding up evolution by facilitating the search for complexity. While arbitrarily complex interaction networks between different modules or molecular species are conceivable, the simplest case applicable to the above system with its two-molecule interactions is the hypercycle~\cite{EigenNaturwissenschaften:78}. Here, $n$ species are arranged in a circular directed graph, where each species enzymatically catalyzes the replication of its next neighbor. This network gives rise to coexistence of all species, in a stable fixed point for $n \leq 4$ and via periodic orbits for larger $n$. In contrast to previous approaches accounting for replication errors in a hypercycle~\cite{CamposPRE:00,SilvestreJTB:08}, we consider distinct error tails for all species: each is present in an enzymatically active variant $x_{\text{e},i}$ with replication rate $r_{\text{e},i} = \alpha_i + \gamma_ix_{\text{e},i+1}$ together with its non-enzymatic error tail $x_{\text{n},i}$ with replication rate $r_{\text{n},i}=\alpha_i$. In addition, there is the global error tail of misfolding mutants. For simplicity, we assume a symmetric setup with identical rate constants $\alpha_i\equiv \alpha$ and $\gamma_i\equiv \gamma$. This gives the rate equations
\begin{equation}\label{eq:rate-equations-hypercycle}
\begin{split}
\dot x_{\text{e},i} &= r_{\text{e},i} Q x_{\text{e},i} - x_{\text{e},i} \bar r \\
\dot x_{\text{n},i} &= r_{\text{n},i} Q^{1-\lambda} x_{\text{n},i} + r_{\text{e},i} (Q^{1-\lambda}-Q)x_{\text{e},i} - x_{\text{n},i} \bar r,
\end{split}
\end{equation}
where indices are taken modulo $n$ and the mean fitness is now given by $\bar r = (\alpha-1)\sum_i (x_{\text{e},i}+x_{\text{n},i}) +\gamma\sum_i x_{\text{e},i}x_{\text{e},i+1}+1$. It is easy to see that this system reduces to Eq.~\eqref{eq:rate-equations-error-tail} if we assume that $x_{\text{e},i}\equiv \xe^*/n$ and $x_{\text{n},i}\equiv \xn^*/n$ and replace $\gamma\to\gamma/n$. Then, the ``enzymatic'' solution of Eq.~\eqref{eq:rate-equations-error-tail} corresponds to the inner fixed point of the hypercycle, where all species are present in equal concentration. Replication errors can be tolerated only if $\mu < \mu_\text{c,e}$, where $\mu_\text{c,e}\sim \ln(\gamma/n)/(2 L)$ for large $\gamma$ (see also Ref.~\cite{CamposPRE:00}). We emphasize that in contrast to Ref.~\cite{ManrubiaRNA:07}, where unspecific ligation was assumed to link functional motifs, in our model specific recognition between different species leads to frequency-dependent replication rates. This reduces the error threshold by roughly a factor of $2$, which in a two-member hypercycle would cancel the putative complexity gain resulting from using two subunits of half the sequence length.

Moreover, increasing the number of hypercycle members beyond $n=4$ changes the stability of the central fixed point. Observing that the Jacobian matrix of Eq.~\eqref{eq:rate-equations-hypercycle} is block-circulant~\cite{TeeRLMS:05} (every block is a $2\times 2$-matrix for the two concentration variables $x_{\text{e},i}$ and $x_{\text{n},i}$ per species), its crucial eigenvalues with possibly non-negative real part are given by $\frac{1}{n}\gamma Q x_\text{e}^* \e^{\imi m\pi/n}$, where $m=0,\ldots,n-1$ and $x_\text{e}^*$ is the enzymatic solution of Eq.~\eqref{eq:rate-equations-error-tail} with $\gamma\to\gamma/n$. Hence, these eigenvalues are proportional to the $n$ different $n$th roots of unity. In close correspondence to the error-free hypercycle (and in contrast to Ref.~\cite{CamposPRE:00}, where a stability region for $n=5$ was found), the central fixed point loses stability for $n > 4$, giving rise to limit cycles with large concentration oscillations, which are vulnerable to extinction via stochastic fluctuations.

Note that the stable inner fixed point corresponding to the enzymatic regime implies coexistence of different species. However, in the non-enzymatic regime of Eq.~\eqref{eq:rate-equations-error-tail}, the different error tails do compete against each other. As soon as the hypercycle breaks down, e.g., because the recognition sequence is lost due to stochastic fluctuations, one error tail will drive the others to extinction, due to the competitive exclusion principle encountered in usual quasispecies theory~\cite{EigenACP:89}. This makes the reverse process, i.e., a fluctuation that establishes a closed cycle, extremly unlikely.

\section{Conclusion}
In summary, we have analyzed a simple quasispecies model for the non-enzymatic and enzymatic replication of ribozymes, where specific recognition is mediated via otherwise neutral sites. We find that the frequency-dependent replication rates associated with specific enzymatic replication lead to a discontinuous transition at the error threshold due to bistability with a partly delocalized phase. Further, hypercyclic couplings enable coexistence of at most four different species and their respective error tails in a stable fixed point.

\acknowledgments
Financial support by the Deutsche Forschungsgemeinschaft through SFB TR12 is gratefully acknowledged.


\end{document}